\pdfoutput=1

\newcommand*{\ATLASLATEXPATH}{}

\documentclass[USenglish,texlive=2016,cernpreprint]{\ATLASLATEXPATH atlasdoc}

\usepackage[subfigure=true]{\ATLASLATEXPATH atlaspackage}
\usepackage{\ATLASLATEXPATH atlasbiblatex}
\usepackage{placeins}
\usepackage{\ATLASLATEXPATH atlasphysics}

\usepackage{ZHccPaper-defs}
\usepackage{multirow}
\usepackage{pdflscape}
\usepackage{verbatim}
\usepackage{amsfonts}
\usepackage{lipsum}
\usepackage{siunitx}

\AtlasTitle{Search for the Decay of the Higgs Boson to Charm Quarks with the ATLAS Experiment}

\AtlasAbstract{%
A direct search for the Standard Model Higgs boson decaying to a pair of charm quarks is presented. Associated production of the Higgs and $Z$ bosons, in the decay mode $ZH\rightarrow \ell^+ \ell^- c \bar{c}$ is studied. A dataset with an integrated luminosity of 36.1 \ifb\ of $pp$ collisions at $\rts=13\,\TeV$ recorded by the ATLAS experiment at the LHC is used. The $H\rightarrow c\bar{c}$ signature is identified using charm-tagging algorithms. The observed (expected) upper limit on $\sigma(pp \rightarrow ZH) \times \mathcal{B}(H \rightarrow c\bar{c})$ is 2.7 ($3.9^{+2.1}_{-1.1}$)~pb at the 95\% confidence level for a Higgs boson mass of 125~GeV, while the Standard Model value is 26~fb.
}

\author{The ATLAS Collaboration}

\AtlasRefCode{HIGG-2017-01}
\AtlasNote{HIGG-2017-01}

\PreprintIdNumber{CERN-EP-2017-334}

\AtlasDate{\today}

\AtlasJournalRef{Phys. Rev. Lett. 120 (2018) 211802}
\AtlasDOI{DOI:10.1103/PhysRevLett.120.211802}

\hypersetup{pdftitle={ATLAS document},pdfauthor={The ATLAS Collaboration}}

\begin{document}

\maketitle

In July 2012, the ATLAS and CMS collaborations announced the discovery of a new particle with a mass of approximately 125~GeV~\cite{HIGG-2012-27,Chatrchyan:2013lba} in searches for the Standard Model (SM) Higgs boson at the Large Hadron Collider (LHC)~\cite{Evans:2008zzb}. Subsequent measurements indicate that this particle is consistent with the SM Higgs boson~\cite{HIGG-2014-06,CMS-HIG-14-009,CMS-HIG-12-041,HIGG-2013-01,CMS-HIG-14-018,HIGG-2015-07,HIGG-2014-14}. Direct evidence for the Yukawa coupling of the Higgs boson to the top~\cite{Aaboud:2017jvq} and bottom~\cite{HIGG-2016-29,Sirunyan:2017elk} quarks was recently obtained. Measurements of the Yukawa coupling of the Higgs boson to quarks in generations other than the third are difficult at hadron colliders, due to small branching fractions, large backgrounds, and challenges in jet flavor identification~\cite{Perez:2015aoa,Perez:2015lra}. This Letter presents a direct search by the ATLAS experiment for the decay of the Higgs boson to a pair of charm ($c$) quarks. This search targets the production of the Higgs boson in association with a $Z$ boson decaying to charged leptons: $Z(\ell^{+}\ell^{-})H(c\bar{c})$, where $\ell=e,\mu$.

The SM branching fraction for a Higgs boson with a mass of $125\,\GeV$ to decay to a pair of charm quarks is predicted to be 2.9\%~\cite{Djouadi:1997yw}. The inclusive cross-section for $\sigma(\pp \rightarrow ZH) \times \mathcal{B}(H\rightarrow \ccbar)$ is 26~fb at $\sqrt{s}=13\,\TeV$~\cite{deFlorian:2016spz}. Rare exclusive decays of the Higgs boson to a light vector meson or quarkonium state and a photon can also probe the couplings of the second-generation quarks to the Higgs boson~\cite{Bodwin:2013gca,Kagan:2014ila,Aad:2015sda,HIGG-2016-05}. Previously, the ATLAS Collaboration presented an indirect search for the decay of the Higgs boson to $c$-quarks via the decay to $\jpsi\gamma$, obtaining a branching fraction limit of $1.5 \times 10^{-3}$ at the 95\% confidence level (CL), which approximately corresponds to a limit of $540$ times the SM prediction~\cite{Aad:2015sda,Perez:2015aoa}. Bounds on Higgs boson branching fractions to unobserved final states and fits to global rates constrain $\mathcal{B}(H \rightarrow \ccbar) < 20$\% at the 95\% CL, assuming SM production cross-sections~\cite{Delaunay:2013pja}. These limits can still accommodate large modifications to the Higgs boson coupling to charm quarks from new physics~\cite{Delaunay:2013pja}. In this Letter, a new approach is introduced to investigate the coupling of the Higgs boson to charm quarks.

The search is performed using $pp$ collision data recorded in 2015 and 2016 with the ATLAS detector~\cite{Aad:2008zzm} at $\sqrt{s}=13$~TeV. The ATLAS detector at the LHC covers nearly the entire solid angle around the collision point.\footnote{ATLAS uses a right-handed coordinate system with its origin at the nominal interaction point (IP)
in the center of the detector and the $z$-axis along the beam pipe.
The $x$-axis points from the IP to the center of the LHC ring,
and the $y$-axis points upwards.
Cylindrical coordinates $(r,\phi)$ are used in the transverse plane,
$\phi$ being the azimuthal angle around the $z$-axis.
The pseudorapidity is defined in terms of the polar angle $\theta$ as $\eta = -\ln \tan(\theta/2)$.
Angular distance is measured in units of $\Delta R \equiv \sqrt{(\Delta\eta)^{2} + (\Delta\phi)^{2}}$.} It consists of an inner tracking detector surrounded by a thin superconducting solenoid, electromagnetic and hadronic calorimeters, and a muon spectrometer incorporating three large superconducting toroidal magnets. An additional pixel layer was installed for the $\sqrt{s} = 13$~TeV running period~\cite{ATLAS-TDR-19}.
After the application of beam, detector and data-quality requirements, the integrated luminosity corresponds to $36.1 \pm 0.8$~fb$^{-1}$, measured following Ref.~\cite{DAPR-2013-01}. Events are required to contain exactly two same-flavor leptons with an invariant mass consistent with that of the $Z$ boson, and at least two jets of which one or two are identified as charm jets ($c$-jets). In this Letter, lepton refers to only electrons or muons. The analysis procedure is validated by measuring the yield of $ZW$ and $ZZ$ production, where the sample is enriched in $W\rightarrow cs, cd$ and $Z\rightarrow \ccbar$ decays. Further details can be found in Ref.~\cite{HIGG-2016-29}.

Monte Carlo (MC) simulated samples were produced for signal and background processes using the full ATLAS detector simulation~\cite{SOFT-2010-01} using \textsc{Geant4}~\cite{Agostinelli:2002hh}. Table~\ref{tbl:evgen} provides details of the event generators used for each signal and background sample. Signal events were produced at next-to-leading order (NLO) for the $q\bar{q} \rightarrow ZH$ process and at leading order (LO) for the $gg \rightarrow ZH$ process with \textsc{Powheg-BOX} v2~\cite{Alioli:2010xd}. The dominant $Z$+jets background and the resonant diboson $ZW$ and $ZZ$ processes were generated using \textsc{Sherpa} 2.2.1~\cite{sherpa}. The \ttbar background was generated using \textsc{Powheg-BOX} v2. Backgrounds from single top and multijet production and the contribution from Higgs decays other than $b\bar{b}$ and $c\bar{c}$ are assessed to be negligible and not considered further. The Higgs boson mass is set to $m_{H} = 125$~GeV and the top-quark mass is set to 172.5 GeV.

\begin{table}
\begin{center}
\caption{\label{tbl:evgen} The configurations used for event generation of the signal and background processes. If two parton distribution functions (PDFs) are shown, the first is for the matrix element calculation and the second for the parton shower, otherwise the same is used for both. Alternative event generators and configurations, used to estimate systematic uncertainties, are in parentheses. Tune refers to the underlying-event tuned parameters of the parton shower event generator. \textsc{MG5\_aMC} refers to \textsc{MadGraph5\_aMC@NLO} 2.2.2~\cite{Alwall:2014hca}; \textsc{Pythia} 8 refers to version 8.212~\cite{Sjostrand:2007gs}. Heavy-flavor hadron decays modeled by \textsc{EvtGen} 1.2.0~\cite{Lange:2001uf} are used for all samples except those generated using \textsc{Sherpa}. The order of the calculation of the cross-sections used to normalize the predictions
is indicated. The $q\bar{q}\rightarrow ZH$ cross-section is estimated by subtracting the $gg\rightarrow ZH$ cross-section from the $pp \rightarrow ZH$ cross-section. The asterisk (*) in the last column denotes that the indicated order is for the $pp \rightarrow ZH$ cross-section. NNLO denotes next-to-next-to-leading order; NLL denotes next-to-leading-log and NNLL denotes next-to-next-to-leading log.}
{\footnotesize
\setlength\tabcolsep{1.5pt}
\begin{tabular}{lllllll}
\hline\hline
Process & Event Generator & Parton Shower & PDF & Tune & Cross-section\\
& (alternative) & (alternative) & (alternative) & \\
\hline
$q\bar{q} \rightarrow ZH$ & \textsc{Powheg-BOX} v2~\cite{Alioli:2010xd} & \textsc{Pythia} 8 & PDF4LHC15NLO~\cite{Butterworth:2015oua} & AZNLO~\cite{STDM-2012-23} & NNLO (QCD)*\\
&+\textsc{GoSaM}~\cite{Cullen:2011ac}& & /CTEQ6L1~\cite{Pumplin:2002vw,Nadolsky:2008zw} & & +NLO~(EW)~\cite{Ciccolini:2003jy,Brein:2003wg,Ferrera:2011bk,Brein:2011vx,Ferrera:2013yga,Ferrera:2014lca,Campbell:2016jau} \\
& +\textsc{MiNLO}~\cite{Hamilton:2012np,Luisoni:2013kna} & (\textsc{Herwig 7}~\cite{Bellm:2015jjp}) && (A14~\cite{ATL-PHYS-PUB-2014-021})& \\
$gg\rightarrow ZH$ & \textsc{Powheg-BOX} v2 & \textsc{Pythia} 8 & PDF4LHC15NLO & AZNLO & NLO+NLL~(QCD)~\cite{Altenkamp:2012sx,Hespel:2015zea,Altenkamp:2012sxAllAuthors,deFlorian:2016spz}\\
& & (\textsc{Herwig 7}) & /CTEQ6L1 & (A14)& \\
$\ttbar$ & \textsc{Powheg-BOX} v2 & \textsc{Pythia} 8 & NNPDF3.0NLO~\cite{Ball:2012cx} & A14 & NNLO+NNLL~\cite{Czakon:2011xx} \\
& & (\textsc{Herwig} 7) & /NNPDF2.3LO & \\
$ZW$, $ZZ$ & \textsc{Sherpa} 2.2.1 \cite{sherpa} & \textsc{Sherpa} & NNPDF3.0NNLO & \textsc{Sherpa} & NLO \\
& (\textsc{Powheg-BOX}) & (\textsc{Pythia} 8) & & \\
$Z$+jets & \textsc{Sherpa} 2.2.1 & \textsc{Sherpa} & NNPDF3.0NNLO & \textsc{Sherpa} &NNLO~\cite{Catani:2009sm}\\
& (\textsc{MG5\_aMC}) & (\textsc{Pythia} 8) & (NNPDF2.3LO) & (A14) & \\
\hline\hline
\end{tabular}
}
\end{center}
\end{table}

Events are required to have at least one reconstructed primary vertex. Electron candidates are reconstructed from energy clusters in the electromagnetic calorimeter that are associated with charged-particle tracks reconstructed in the inner detector~\cite{PERF-2016-01,ATLAS-CONF-2016-024}. Muon candidates are reconstructed by combining inner detector tracks with
muon spectrometer tracks or energy deposits in the calorimeters consistent with the passage of minimum-ionizing particles~\cite{Aad:2016jkr}. For data recorded in 2015, the single-electron (muon) trigger required a candidate with $\pT > 24~(20)$~\GeV; in 2016 the lepton \pT threshold was raised to 26 GeV. Events are required to contain a pair of same-flavor leptons, both satisfying $\pT > 7$~GeV and $|\eta|< 2.5$. At least one lepton must have $\pT>27$~GeV and correspond to a lepton that passed the trigger. The two leptons are required to satisfy loose track-isolation criteria with an efficiency greater than 99\%. They are required to have opposite charge in dimuon events, but not in dielectron events due to the non-negligible charge misidentification rate of electrons. The invariant mass of the dilepton system is required to be consistent with the mass of the $Z$ boson: $81~\mathrm{GeV} < m_{\ell\ell} < 101$~GeV.

Jets are reconstructed from topological energy clusters in the calorimeters~\cite{PERF-2014-07,ATL-PHYS-PUB-2015-036} using the \antikt algorithm~\cite{AntiKt} with a radius parameter of 0.4 implemented in the \textsc{FastJet} package~\cite{Cacciari:2011ma}. The jet energy is corrected using a jet-area-based technique~\cite{Cacciari:2007fd,PERF-2014-03} and calibrated~\cite{PERF-2011-03,PERF-2016-04} using \pT- and $\eta$-dependent correction factors determined from simulation, with residual corrections from internal jet properties. Further corrections from in situ measurements are applied to data. Selected jets must have $\pT > 20$~GeV and $|\eta| < 2.5$. Events are required to contain at least two jets. If a muon is found within a jet, its momentum is added to the selected jet. An overlap removal procedure resolves cases in which the same physical object is reconstructed multiple times, e.g. an electron also reconstructed as a jet.

Jets in simulated events are labeled according to the presence of a heavy-flavor hadron with $\pT > 5$~GeV within $\Delta R = 0.3$ from the jet axis. If a $b$-hadron is found the jet is labeled as a $b$-jet. If no $b$-hadron is found, but a $c$-hadron is present, then the jet is labeled as a $c$-jet. Otherwise the jet is labeled as a light-flavor jet ($l$-jet).

\begin{figure}
\centering
\includegraphics[width=0.5\textwidth]{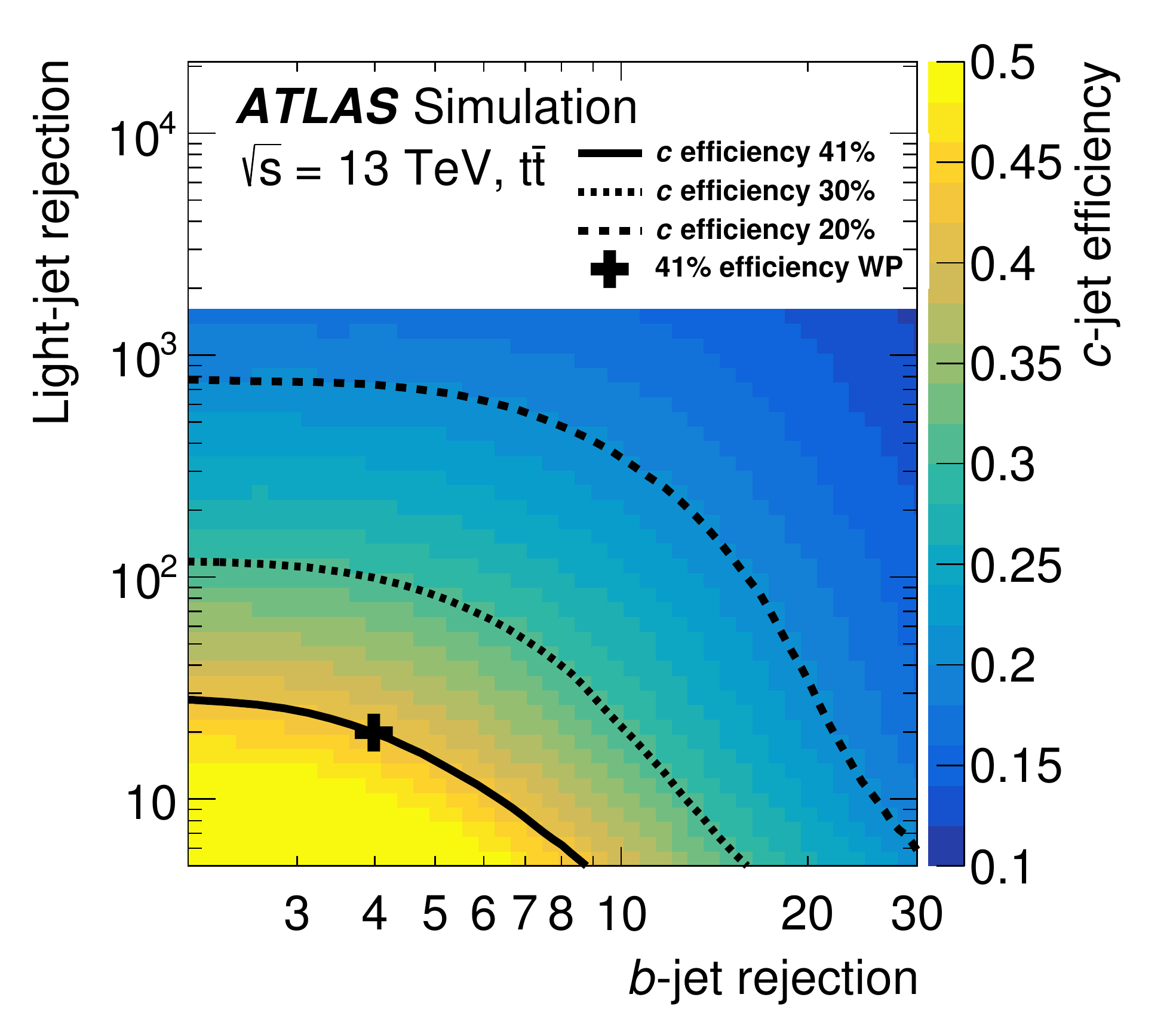}
\caption{The $c$-jet tagging efficiency (colored scale) as a function of the $b$-jet and $l$-jet rejection as obtained from simulated \ttbar events. The cross, labeled as working point, WP, denotes the selection criterion used in this analysis. The solid and dotted black lines indicate the contours in rejection space for the fixed $c$-tagging efficiency used in the analysis and two alternatives.}
\label{fig:ctagging}
\end{figure}

Flavor-tagging algorithms exploit the different lifetimes of $b$-, $c$- and light-flavor hadrons. A $c$-tagging algorithm is used to identify $c$-jets. Charm jets are particularly challenging to tag because $c$-hadrons have shorter lifetimes and decay to fewer charged particles than $b$-hadrons. Boosted decision trees (BDTs) are trained to obtain two multivariate discriminants: to separate \cjets from \ljets and \cjets from \bjets. The same variables used for $b$-tagging~\cite{PERF-2012-04,ATL-PHYS-PUB-2016-012} are used. Figure~\ref{fig:ctagging} shows the selection criteria applied in the two-dimensional multivariate discriminant space, to obtain an efficiency of 41\% for \cjets and rejection factors of 4.0 and 20 for \bjets and \ljets. The efficiencies are calibrated to data using $b$-quarks from $t \rightarrow Wb$ and $c$-quarks from $W \rightarrow cs,cd$ with methods identical to the $b$-tagging algorithms~\cite{PERF-2012-04}. Statistical uncertainties in the simulation are reduced, by weighting events according to the tagging efficiencies of their jets, parameterized as a function of jet flavor, \pT, $\eta$ and the angular separation between jets, rather than imposing a direct requirement on the $c$-tagging discriminants.

Data are analyzed in four categories with different expected signal purities. The dijet invariant mass, $m_{\ccbar}$, constructed using the two highest-\pT jets, is the discriminating variable in each category. Categories are defined using the transverse momentum of the reconstructed $Z$ boson, \ptZ~ ($75~\GeV \leq \ptZ < 150$~\GeV~and $\ptZ \geq 150$~\GeV)~and the number of $c$-tags amongst the leading jets (either one or two). The $\ptZ$ requirements exploit the harder $\ptZ$ distribution in $ZH$ compared to $Z+\text{jets}$ production. Background events are rejected by requiring the angular separation between the two jets constituting the dijet system, $\Delta R_{c\cbar}$, to be less than 2.2, 1.5 or 1.3 for events satisfying $75 \leq \ptZ < 150$~GeV, $150 \leq \ptZ < 200$~GeV or $\ptZ \geq 200$~GeV. The signal acceptance ranges from 0.5\% to 3.4\% depending on the category. A joint binned maximum-profile-likelihood fit to $m_{c\cbar}$ in the categories is used to extract the signal yield and the $Z$+jets background normalization. The fit uses 15 bins in each category within the range of $50~\GeV < \mcc < 200$~\GeV, with a bin width of $10\,\GeV$. The parameter of interest, $\mu$, common to all categories, is the signal strength, defined as the ratio of the measured signal yield to the SM prediction.

Systematic uncertainties affecting the signal and background predictions include theoretical uncertainties in the signal and background modeling and experimental uncertainties. Table~\ref{tab:systs} shows their relative impact on the fitted value of $\mu$. Uncertainties in the \mcc shape of the backgrounds are assessed by comparisons between nominal and alternative event generators as indicated in Table~\ref{tbl:evgen}.

Systematic uncertainties are incorporated within the statistical model through nuisance parameters that modify the shape and/or normalization of the distributions. Statistical uncertainties in the simulation samples are accounted for. The $Z$+jets background is normalized from the data through the inclusion of an unconstrained normalization parameter for each category. The fitted normalization parameters range between 1.13 and 1.30. All other background normalization factors are correlated between categories, with acceptance uncertainties of order 10\% to account for relative variations between categories.

The dominant contributions to the uncertainty in $\mu$ are the efficiency of the tagging algorithms, the jet energy scale and resolution, and the background modeling. The largest uncertainty is due to the normalization of the dominant $Z$+jets background. The typical uncertainty in the tagging efficiency is 25\% for $c$-jets, 5\% for $b$-jets, and 20\% for $l$-jets.

\begin{table}[htbp]
\caption{Breakdown of the relative contributions to the total uncertainty in $\mu$. The statistical uncertainty includes the contribution from the floating $Z$+jets normalization parameters. The sum in quadrature of the individual components differs from the total uncertainty due to correlations between the components.
~\label{tab:systs}}
\begin{center}
\begin{tabular}{ l | r }
\hline\hline
Source & $\sigma/\sigma_{\mathrm{tot}}$ \\
\hline
\textbf{Statistical} & \textbf{49\%} \\
 \ \ Floating $Z+\text{jets}$ normalization & 31\% \\
\textbf{Systematic} & \textbf{87\%} \\
 \ \ Flavor tagging & 73\% \\
 \ \ Background modeling & 47\% \\
 \ \ Lepton, jet and luminosity & 28\% \\
 \ \ Signal modeling & 28\% \\
 \ \ MC statistical & 6\% \\
\hline\hline
\end{tabular}
\end{center}
\end{table}

\begin{table}[htbp]
\caption{Post-fit yields for the signal and background processes in each category from the profile likelihood fit. Uncertainties include statistical and systematic contributions. The pre-fit SM expected $ZH(c\bar{c})$ signal yields are indicated in parenthesis. ~\label{tab:yields}}
\small
\begin{center}
\scalebox{0.9}{
{\renewcommand{\arraystretch}{1.2}
\begin{tabular}{l | r@{$\,\pm\,$}l | r@{$\,\pm\,$}l | r@{$\,\pm\,$}l | r@{$\,\pm\,$}l }
\hline\hline
\multirow{2}{*}{Sample} & \multicolumn{8}{c}{Yield, $50~\GeV < \mcc < 200~\GeV$} \\
\cline{2-9}
& \multicolumn{4}{c|}{1 $c$-tag} & \multicolumn{4}{c}{2 $c$-tags} \\
\hline
& \multicolumn{2}{c|}{$75 \leq p_{\mathrm{T}}^{Z} < 150\,\GeV$} & \multicolumn{2}{c|}{$p_{\mathrm{T}}^{Z} \geq 150\,\GeV$} & \multicolumn{2}{c|}{$75 \leq p_{\mathrm{T}}^{Z} < 150\,\GeV$} & \multicolumn{2}{c}{$p_{\mathrm{T}}^{Z} \geq 150\,\GeV$} \\
\hline
$Z+\text{jets}$ & 69400 & 500 & 15650 & 180 & 5320 & 100 & 1280 & 40 \\
$ZW$ & 750 & 130 & 290 & 50 & 53 & 13 & 20 & 5 \\
$ZZ$ & 490 & 70 & 180 & 28 & 55 & 18 & 26 & 8 \\
$\ttbar$ & 2020 & 280 & 130 & 50 & 240 & 40 & 13 & 6 \\
$ZH(\bbbar)$ & 32 & 2 & 19.5 & 1.5 & 4.1 & 0.4 & 2.7 & 0.2 \\
$ZH(\ccbar)$ (SM) & -143 & 170 (2.4) & -84 & 100 (1.4) & -30 & 40 (0.7) & -20 & 29 (0.5) \\
\hline
Total & 72500 & 320 & 16180 & 140 & 5650 & 80 & 1320 & 40 \\
\hline
Data & \multicolumn{2}{l|}{72504} & \multicolumn{2}{l|}{16181} & \multicolumn{2}{l|}{5648} & \multicolumn{2}{l}{1320} \\
\hline\hline
\end{tabular}
}
}
\end{center}
\end{table}

\begin{figure}[h!]
\centering
\subfigure[]{\includegraphics[width=0.42\textwidth]{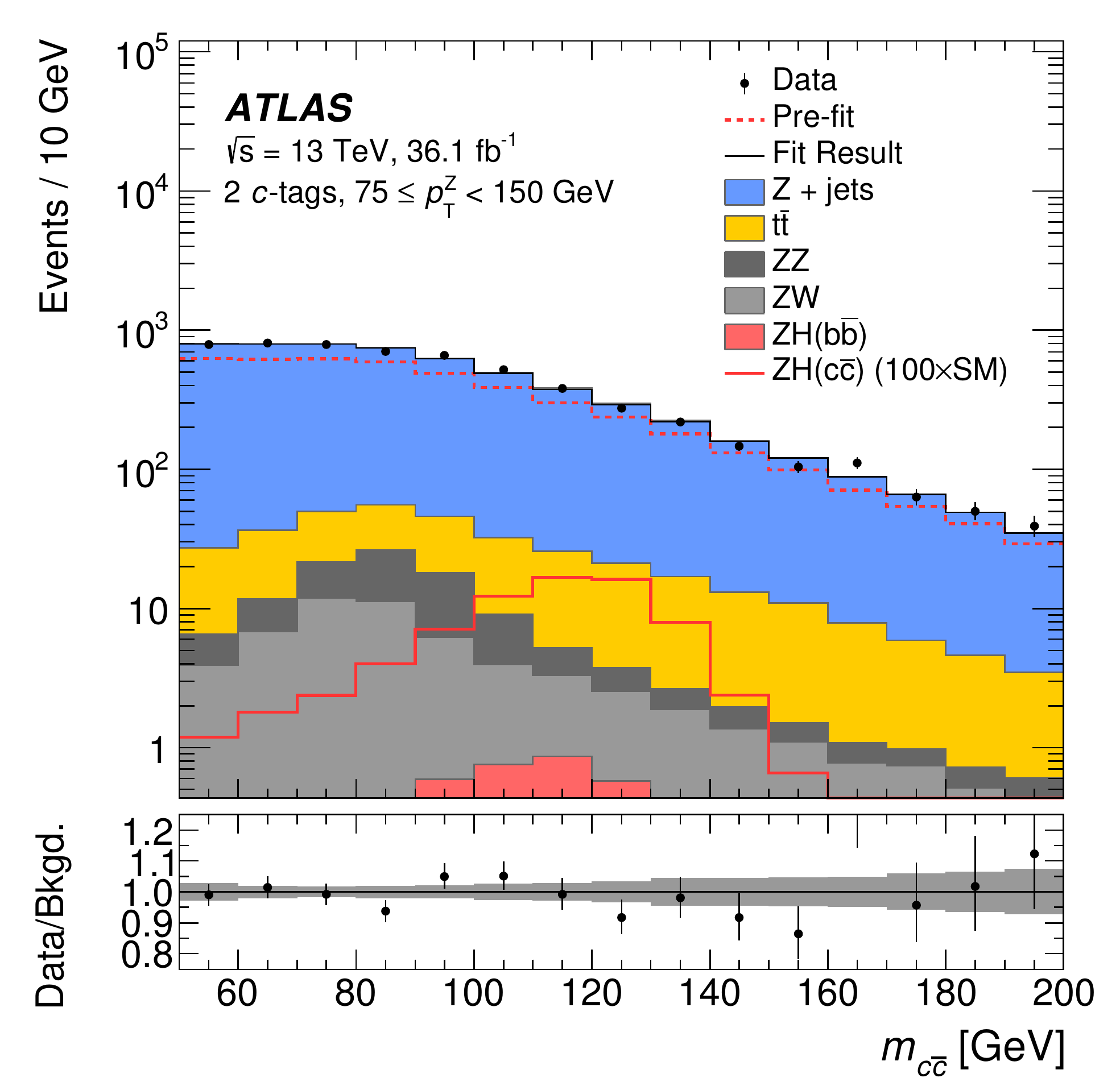}} 
\subfigure[]{\includegraphics[width=0.42\textwidth]{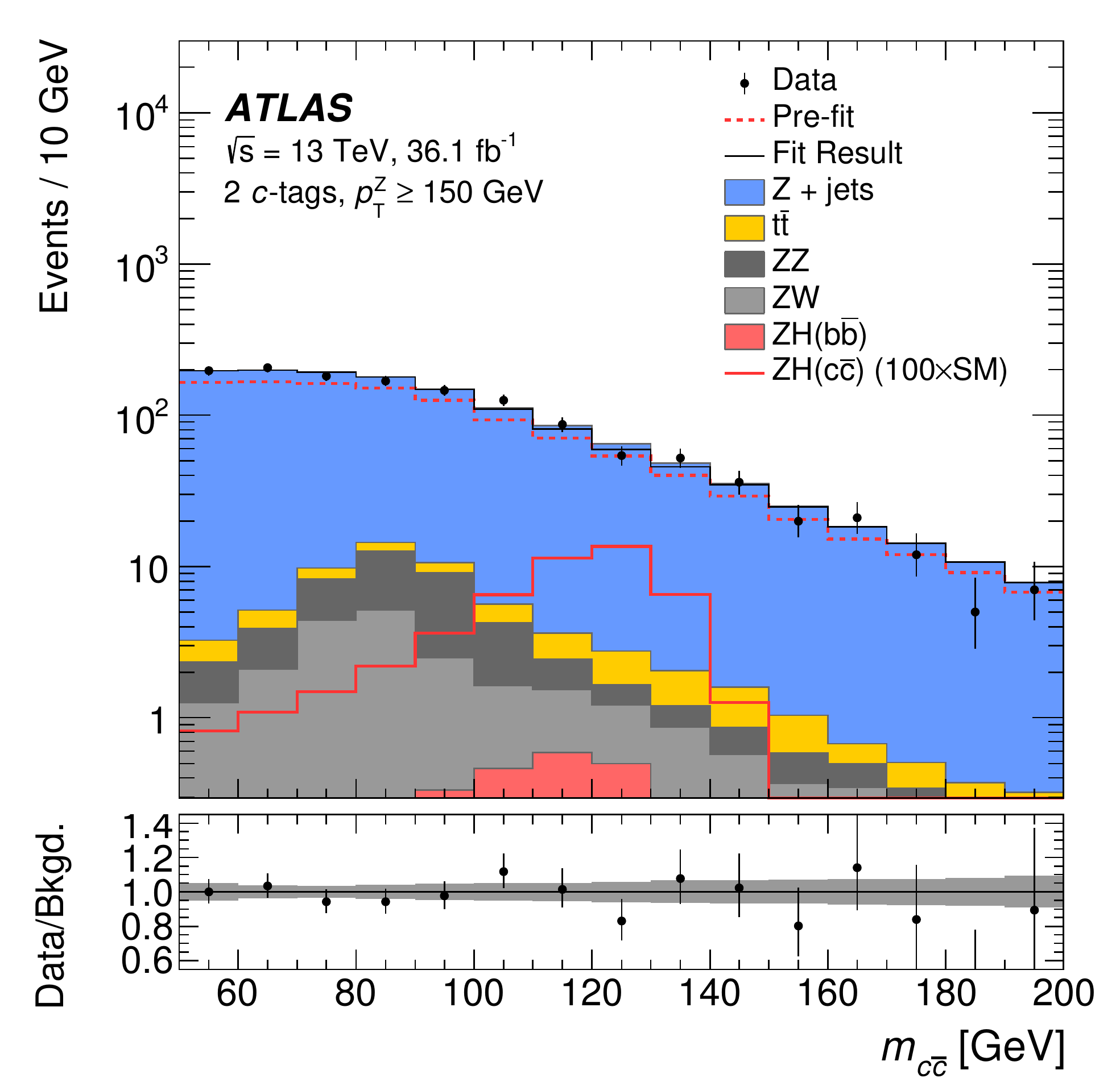}} 
\caption{Observed and predicted $m_{c\cbar}$ distributions in the 2 $c$-tag analysis categories. The expected signal is scaled by a factor of 100. Backgrounds are corrected to the results of the fit to the data. The predicted background from the simulation is shown as red dashed histograms. The ratios of the data to the fitted background are shown in the lower panels. The error bands indicate the sum in quadrature of the statistical and systematic uncertainties in the background prediction.} 
\label{fig:fit_result}
\end{figure}

Table~\ref{tab:yields} shows the fitted signal and background yields. The $m_{c\cbar}$ distributions in the 2 $c$-tag categories are shown in Figure~\ref{fig:fit_result} with the background shapes and normalizations according to the result of the fit. Good agreement is observed between the post-fit shapes of the distributions and the data.

The analysis procedure is validated by measuring the yield of $ZV$ production, where $V$ denotes a $W$ or $Z$ boson, with the same event selection.
The fraction of the $ZZ$ yield from $Z \rightarrow c\bar{c}$ decays is $\sim55\%$ ($20\%$) in the 2 $c$-tag (1 $c$-tag) category, while the fraction of the $ZW$ yield from $W \rightarrow cs,cd$ is $\sim65\%$ for both the 2 and 1 $c$-tag categories.
Contributions of Higgs boson decays to $c\cbar$ and $b\bbar$ are treated as background and constrained to the SM predictions within its theoretical uncertainties. The diboson signal strength is measured to be $\mu_{ZV} = 0.6^{+0.5}_{-0.4}$ with an observed (expected) significance of 1.4 ($2.2$) standard deviations.

The best-fit value for the $ZH(\ccbar)$ signal strength is $\mu_{ZH}=-69 \pm 101$. By assuming a signal with the kinematics of the SM Higgs boson, model-dependent corrections are made to extrapolate to the inclusive phase space. Hence, an upper limit on $\sigma(pp \rightarrow ZH) \times \mathcal{B}(H \rightarrow c\bar{c})$ is computed using a modified frequentist CL$_\text{s}$ method~\cite{Cowan:2010js,Read:2002hq} with the profile likelihood ratio as the test statistic. The observed (expected) upper limit is found to be 2.7 ($3.9^{+2.1}_{-1.1}$)~pb at the 95\% CL. This corresponds to an observed (expected) upper limit on $\mu$ at the 95\% CL of 110 ($150^{+80}_{-40}$). The uncertainties in the expected limits correspond to the $\pm1\sigma$ interval of background-only pseudo-experiments. With the current sensitivity, the result depends weakly on the assumption of the SM rate for $H \rightarrow b\bar{b}$. The observed limit remains within 5\% of the nominal value when the assumed value for normalization of the $ZH(b\bar{b})$ background is varied from zero to twice the SM prediction.

\FloatBarrier

A search for the decay of the Higgs boson to charm quarks has been performed using 36.1~fb$^{-1}$ of data collected with the ATLAS detector in $pp$ collisions at $\sqrt{s}=13$~TeV at the LHC. No significant excess of $ZH(\ccbar)$ production is observed over the SM background expectation. The observed upper limit on $\sigma(pp \rightarrow ZH) \times \mathcal{B}(H \rightarrow c\bar{c})$ is 2.7~pb at the 95\% CL. The corresponding expected upper limit is $3.9^{+2.1}_{-1.1}\,\mathrm{pb}$. This is the most stringent limit to date in direct searches for the inclusive decay of the Higgs boson to charm quarks.

We thank CERN for the very successful operation of the LHC, as well as the
support staff from our institutions without whom ATLAS could not be
operated efficiently.

We acknowledge the support of ANPCyT, Argentina; YerPhI, Armenia; ARC, Australia; BMWFW and FWF, Austria; ANAS, Azerbaijan; SSTC, Belarus; CNPq and FAPESP, Brazil; NSERC, NRC and CFI, Canada; CERN; CONICYT, Chile; CAS, MOST and NSFC, China; COLCIENCIAS, Colombia; MSMT CR, MPO CR and VSC CR, Czech Republic; DNRF and DNSRC, Denmark; IN2P3-CNRS, CEA-DRF/IRFU, France; SRNSFG, Georgia; BMBF, HGF, and MPG, Germany; GSRT, Greece; RGC, Hong Kong SAR, China; ISF, I-CORE and Benoziyo Center, Israel; INFN, Italy; MEXT and JSPS, Japan; CNRST, Morocco; NWO, Netherlands; RCN, Norway; MNiSW and NCN, Poland; FCT, Portugal; MNE/IFA, Romania; MES of Russia and NRC KI, Russian Federation; JINR; MESTD, Serbia; MSSR, Slovakia; ARRS and MIZ\v{S}, Slovenia; DST/NRF, South Africa; MINECO, Spain; SRC and Wallenberg Foundation, Sweden; SERI, SNSF and Cantons of Bern and Geneva, Switzerland; MOST, Taiwan; TAEK, Turkey; STFC, United Kingdom; DOE and NSF, United States of America. In addition, individual groups and members have received support from BCKDF, the Canada Council, CANARIE, CRC, Compute Canada, FQRNT, and the Ontario Innovation Trust, Canada; EPLANET, ERC, ERDF, FP7, Horizon 2020 and Marie Sk{\l}odowska-Curie Actions, European Union; Investissements d'Avenir Labex and Idex, ANR, R{\'e}gion Auvergne and Fondation Partager le Savoir, France; DFG and AvH Foundation, Germany; Herakleitos, Thales and Aristeia programmes co-financed by EU-ESF and the Greek NSRF; BSF, GIF and Minerva, Israel; BRF, Norway; CERCA Programme Generalitat de Catalunya, Generalitat Valenciana, Spain; the Royal Society and Leverhulme Trust, United Kingdom.

The crucial computing support from all WLCG partners is acknowledged gratefully, in particular from CERN, the ATLAS Tier-1 facilities at TRIUMF (Canada), NDGF (Denmark, Norway, Sweden), CC-IN2P3 (France), KIT/GridKA (Germany), INFN-CNAF (Italy), NL-T1 (Netherlands), PIC (Spain), ASGC (Taiwan), RAL (UK) and BNL (USA), the Tier-2 facilities worldwide and large non-WLCG resource providers. Major contributors of computing resources are listed in Ref.~\cite{ATL-GEN-PUB-2016-002}.

\printbibliography

\clearpage
\begin{flushleft}
{\Large The ATLAS Collaboration}

\bigskip

M.~Aaboud$^\textrm{\scriptsize 137d}$,
G.~Aad$^\textrm{\scriptsize 88}$,
B.~Abbott$^\textrm{\scriptsize 115}$,
O.~Abdinov$^\textrm{\scriptsize 12}$$^{,*}$,
B.~Abeloos$^\textrm{\scriptsize 119}$,
S.H.~Abidi$^\textrm{\scriptsize 161}$,
O.S.~AbouZeid$^\textrm{\scriptsize 139}$,
N.L.~Abraham$^\textrm{\scriptsize 151}$,
H.~Abramowicz$^\textrm{\scriptsize 155}$,
H.~Abreu$^\textrm{\scriptsize 154}$
% ...
\bigskip
\\
% ...
$^{12}$ Institute of Physics, Azerbaijan Academy of Sciences, Baku, Azerbaijan\\
% ...
$^{88}$ CPPM, Aix-Marseille Universit{\'e} and CNRS/IN2P3, Marseille, France\\
% ...
$^{115}$ Homer L. Dodge Department of Physics and Astronomy, University of Oklahoma, Norman OK, United States of America\\
% ...
$^{119}$ LAL, Univ. Paris-Sud, CNRS/IN2P3, Universit{\'e} Paris-Saclay, Orsay, France\\
% ...
$^{137}$ $^{(a)}$ Facult{\'e} des Sciences Ain Chock, R{\'e}seau Universitaire de Physique des Hautes Energies - Universit{\'e} Hassan II, Casablanca; $^{(b)}$ Centre National de l'Energie des Sciences Techniques Nucleaires, Rabat; $^{(c)}$ Facult{\'e} des Sciences Semlalia, Universit{\'e} Cadi Ayyad, LPHEA-Marrakech; $^{(d)}$ Facult{\'e} des Sciences, Universit{\'e} Mohamed Premier and LPTPM, Oujda; $^{(e)}$ Facult{\'e} des sciences, Universit{\'e} Mohammed V, Rabat, Morocco\\
% ...
$^{139}$ Santa Cruz Institute for Particle Physics, University of California Santa Cruz, Santa Cruz CA, United States of America\\
% ...
$^{151}$ Department of Physics and Astronomy, University of Sussex, Brighton, United Kingdom\\
% ...
$^{154}$ Department of Physics, Technion: Israel Institute of Technology, Haifa, Israel\\
$^{155}$ Raymond and Beverly Sackler School of Physics and Astronomy, Tel Aviv University, Tel Aviv, Israel\\
% ...
$^{161}$ Department of Physics, University of Toronto, Toronto ON, Canada\\
% ...
$^{a}$ Also at Department of Physics, King's College London, London, United Kingdom\\
$^{b}$ Also at Institute of Physics, Azerbaijan Academy of Sciences, Baku, Azerbaijan\\
$^{c}$ Also at Novosibirsk State University, Novosibirsk, Russia\\
$^{d}$ Also at TRIUMF, Vancouver BC, Canada\\
$^{e}$ Also at Department of Physics {\&} Astronomy, University of Louisville, Louisville, KY, United States of America\\
% ...
$^{*}$ Deceased
\end{flushleft}



\end{document}